\documentstyle[sprocl]{article}

\newcommand{\nc}{\newcommand}
\nc{\be}[1]{\begin{equation} \mbox{$\label{#1}$}}
\nc{\bea}[1]{\begin{eqnarray}\mbox{$\label{#1}$}}
\nc{\ee}{\end{equation}}
\nc{\eea}{\end{eqnarray}}
\nc{\Label}[1]{\label{#1}}
\nc{\bi}{\bibitem}
\nc{\sss}{\scriptscriptstyle}
\nc{\lsim}{\mbox{\raisebox{-.6ex}{~$\stackrel{<}{\sim}$~}}}
\nc{\gsim}{\mbox{\raisebox{-.6ex}{~$\stackrel{<}{\sim}$~}}}
\nc{\nn}{\nonumber}

\def\NBR#1{{\left( #1 \right)}}                  

\def\DBR#1#2{\bigl\{#1 \bigr\}\bigl\{#2 \bigr\}} 

\def\ie{{\em i.e.\ }}

\def\calC{{\cal C}}
\begin{document}
%
%
\title{QUANTUM BOLTZMANN EQUATIONS FOR MIXING SCALAR FIELDS}

\author{M. JOYCE$^{1)}$, K. KAINULAINEN$^{2)}$ and T. PROKOPEC$^{3)}$}

\address{~\\$^1$\it INFN, Sezione di Roma 1, Italy\\
         ~\\$^2$NORDITA, Blegdamsvej 17, DK-2100, Copenhagen \O , Denmark \\
         ~\\$^3$Niels Bohr Institute,
                Blegdamsvej 17, DK-2100, Copenhagen \O , Denmark}

\maketitle\abstracts{
We report on a work in progress, whose goal is a systematic field theoretical
derivation of the quantum transport equations for baryon production in the
electroweak plasma at a first order phase transition in the limit of slowly
varying background fields (thick wall limit).  We start with the
Schwinger-Dyson equations for the two point Green function written in the 
closed time contour (CTC) formalism. The quantum Boltzmann equations for 
the density matrix arise when the SD-equations are expanded to the first 
order in the gradients in the on-shell limit. In this paper we consider only 
scalar fields, but the formalism can easily be extended to fermions. }

%
%
\section{Introduction}
\label{sec:Introduction}
The matter-antimatter asymmetry to be explained is most accurately
pinned down by nucleosynthesis considerations. Allowing for a fair uncertainty
in the present observational situation, one has \cite{keith}
\begin{equation}
\eta_B \equiv \frac{n_B}{s}\simeq 2-7 \times 10^{-11}.
\end{equation}
An exciting realisation is that $\eta_B$ could have been generated at the 
electroweak phase transition. What is required is a strong transition and
an adequate source of CP-violation. While the minimal standard model (MSM)
is known to fail these conditions \cite{kari}, they may well be met in its
popular minimal supersymmetric extension (MSSM) \cite{qj}. There are three
distinct parts of the problem:
(1) {\it equilibrium calculations}, which provide information about the
strength of the phase transition \cite{kari};
(2) {\it phase transition dynamics}, whose study provides the phase boundary
shape and speed \cite{mp}; and finally
(3) {\it baryon production computation}. The goal of our program
is to bring problem (3) to the level of understanding of (1).

At the moment there is no systematic quantum field theoretical derivation of
the CP-violating sources and the transport equations that would consistently
describe all aspects of the problem in one formalism. The difficulty lies in
the inherent out-of-equlibrium nature of the problem. The theory we study
here, is defined by the following scalar lagrangian
\begin{equation}
  {\cal L} =
   \NBR{\partial_\mu \phi}^{\dagger} \NBR{\partial^\mu \phi}
   - \phi^\dagger m^2(x) \phi - {\cal L}_{\rm int}\,,
\label{Lagrange}
\end{equation}
where ${\cal L}_{\rm int}$ contains the interaction terms. The field $\phi$
can have several components so that $m^2$ is in general a complex matrix
varying in space and time, possibly giving rise to out-of-equilibrium
conditions
and nontrivial CP-properties, which are the necessary ingredients for
baryogenesis.

%
%
\section{The Kadanoff-Baym equations}
\label{sec: KB equations}

We start our analysis with the exact Schwinger-Dyson equations in the Keldysh
closed time contour (CTC) formalism \cite{dmr,heinz,henning}.
The basic quantity is the out-of-equilibrium 2-point Green function
\begin{equation}
G^{\alpha\beta}_\calC(x,y) =
           -i \left\langle T_\calC
                 \left[\phi_\alpha (x)\phi^\dagger_\beta (y)\right]
             \right\rangle
\label{Gcontour}
\end{equation}
where $T_\calC$ defines time ordering along the contour $\calC$ which starts at
some $t_0$ (often taken to be at $-\infty$) goes to $+\infty$, and then back to
$t_0$. From now on we suppress the indices $\alpha,\beta$, which denote
different scalar fields (``flavours''), and work with matrices in this
flavour space. The two point function $G_\calC(x,y)$ obeys the contour
Schwinger-Dyson equation 
%
\begin{equation}
G_\calC (x,y) = G^0_\calC (x,y)
            + \int_\calC dx' \int_\calC dx'' \; G^0_\calC (x,x')
              \Sigma_\calC (x',x'') G_\calC (x'',y)\,,
\label{SDI}
\end{equation}
where $\Sigma_\calC$ is the self-energy and $G^0_{\calC}$ is the free particle
(tree-level) propagator. In order to solve $G_\calC(x,y)$ from (\ref{SDI}),
some additional information about $\Sigma_\calC$ must be provided. In general
$\Sigma_\calC$ is specified through the usual BBGKY hierarchy of equations for
$n$-point functions. In the weak coupling limit $\Sigma_\calC$ can be
computed perturbatively, if the interaction lagrangian
${\cal L}_I$ in Eqn.~(\ref{Lagrange}) is given; for the moment
however, we can keep $\Sigma_\calC$ completely general.

%
%
After some algebra Eqn.~(\ref{SDI}) can be recast as
the well-known {\em Kadanoff-Baym} (KB) equations  \cite{kb}:
\begin{eqnarray}
  & (G_0^{-1}-\Sigma ^{r,a})\otimes G^{r,a} = \delta & \nonumber\\
  & (G_0^{-1}-\Sigma ^{r})  \otimes G^{<,>} = \Sigma ^{<,>}\otimes G^{a}. &
\label{KB1b}
\end{eqnarray}
where $\otimes$ denotes the convolution integral: $[f \otimes g](x,y) 
\equiv \int dx' f(x,x')g(x',y)$ and $G^{r,a}$ are the retarded and advanced 
Green functions and 
\begin{eqnarray}
G^>(x,y)       &=& -i \langle \phi(x)\phi^\dagger(y) \rangle
\,,\qquad
G^<(x,y)       = -i \langle \phi^\dagger(y)\phi(x) \rangle
\label{GFs}
\end{eqnarray}
are the quantum Wigner functions. The Kadanoff-Baym equations~(\ref{KB1b}) 
are still exact, but formidable to solve. They considerably simplify when 
expanded to the first order in gradients.

%

\section{Gradient expansion}
\label{sec: Gradient expansion}

Gradient expansion is valid in the limit of slowly varying
background fields. This means that the
Compton wave length $\ell \sim k^{-1}$ of excitations should be
smaller than the characteristic length scale of the background variation.
More precisely, propagators and self-energies should vary slowly with respect
to the macroscopic (average) coordinate $X=(x+y)/2$, \ie
\be{criteria}
|G| \gg |\partial_X\partial_k G| \;, \qquad
|\Sigma| \gg |\partial_X\partial_k \Sigma|\,,
\ee
where the precise meaning of momentum $k$ is defined below.
Applying this {\it gradient expansion} to Eqns.~(\ref{KB1b})
we get the new KB-equations
\begin{eqnarray}
  &e^{-i\Diamond }\DBR{G_0^{-1} - \Sigma^{r,a}}{G^{r,a}}
     = 1 &\label{KB2a} \\
  &e^{-i\Diamond }\DBR{G_0^{-1} - \Sigma^{r}}{G^{<,>}}
     = e^{-i\Diamond}\DBR{\Sigma^{<,>}}{G^{a}}\,,&
\label{KB2b}
\end{eqnarray}
where the Green functions $G=G(k;X)$ and self-energies $\Sigma=\Sigma(k;X)$
are now defined in the mixed representation, defined by the Fourier 
transform with respect to the relative coordinates $r=x-y$
(Wigner transform):
\begin{equation}
  G(k;X) = \int d^4r e^{ik\cdot r} G(X+r/2,X-r/2)\,.
\label{WT}
\end{equation}
The $\Diamond$-operator of Eqn.~(\ref{KB2b}) is defined as the following
generalization of the Poisson brackets:
\begin{equation}
\Diamond\{f\}\{g\} = \frac{1}{2}\left[
                   \partial_X f \cdot \partial_k g
                 - \partial_k f \cdot \partial_X g \right].
\label{diamond}
\end{equation}
To the first order in gradients, $e^{-i\Diamond}$ in Eqn.~(\ref{KB2b})
simplifies to $1-i\Diamond$. For the electroweak phase
boundary the typical macroscopic scale is of the order of the boundary
thickness $L\sim (10-20)/T$ \cite{cjk1,mp}, whereas for a typical plasma
excitation $\partial_k \sim k^{-1} \sim T^{-1}$, so that $\partial_X\partial_k
\sim 1/LT \ll 1$, \ie criteria (\ref{criteria}) are satisfied.

To the lowest order in gradients equations (\ref{KB2a}--\ref{KB2b}) clearly
define the familiar propagators of an excitation in a translationally invariant
background, and their poles define the well known quasiparticle  dispersion
relations. Thus Eqn.~(\ref{KB2a}) in particular contains information
about the spectrum of excitations of the system, and it is contained in the
spectral function defined by
\begin{equation}
{\cal A} \equiv \frac{i}{2}\left( G^r - G^a \right)
              = \frac{i}{2}\left( G^> - G^< \right).
\label{eq: A}
\end{equation}
%

%
%

\section{Propagator equation}
\label{sec: Propagator equation}

We begin by considering the propagator equation for the retarded and advanced
Green functions. For the one-field case Eqns.~(\ref{KB2a}) imply
\begin{eqnarray}
  \cos\Diamond \DBR{G_0^{-1}-\Sigma^{r,a}}{G^{r,a}} &=& 1
\label{spec1}\\
  \sin\Diamond \DBR{G_0^{-1}-\Sigma^{r,a}}{G^{r,a}} &=& 0\,.
\label{spec2}
\end{eqnarray}
Now observe that Eqn.~(\ref{spec2}) can be obtained from Eqn.~(\ref{spec1}) by
an application of the differential operator $\tan\Diamond$, and hence it gives
no new information. Therefore, to the first order in gradients, we have simply
\begin{equation}
G^{r,a} = (G_0^{-1} -\Sigma^{r,a})^{-1}.
\label{Gra: s1}
\end{equation}
From this, one can easily read off the spectral function ${\cal A}$ 
(\ref{eq: A}) and $G_R=(G^r+G^a)/2$.

In the case of several flavours Eqn.~(\ref{Gra: s1}) does not solve the 
propagator equation~(\ref{KB2a}) to the first order in gradients. For 
simplicity consider $\Sigma_R\equiv (\Sigma^r+\Sigma^a)/2\rightarrow 0$. 
Then the explicit solution has the following form
\begin{equation}
G^{r,a}={\cal U}^\dagger\frac{{\rm det}\left[\lambda\right]}
{ {\rm det}\left[\lambda\right]\lambda
  +k_z\bigl[\bigl[H_z,\lambda\bigr]_-,\lambda\bigr]_-
\pm i\epsilon\omega({\rm det}[\lambda]+\lambda{\rm tr}[\lambda])} {\cal U}\,,
\label{Gra: s2}
\end{equation}
where ${\cal U}$ is the unitary matrix in the flavour space that diagonalizes
the mass matrix $m^2$, $H_z=-i{\cal U}\partial_z{\cal U}^\dagger$ is
a ``gauge'' field, and $\lambda={\cal U}(G_0^{-1}-\Sigma_R){\cal U}^\dagger$
is the diagonal matrix with the entries $\lambda_\pm=\omega^2-\omega_\pm^2$,
$\omega_\pm^2=\vec k^2+m_\pm^2$, and $m_\pm^2$ are the  diagonal entries of
$m_D^2={\cal U}m^2{\cal U}^\dagger$ for the two field case. 
Note that an inhomogeneous background alters the pole structure of the
propagator. Indeed, Eqn.~(\ref{Gra: s2}) can be diagonalized, and one can
show that each of the homogeneous particle poles $\omega_\pm$ is
split into three poles, each with a weight 1/3 of the original
excitation, which lie on the circle in the complex frequency
plane, with shifts given by $\{\delta\omega,\delta\omega e^{2i\pi/3},
\delta\omega e^{-2i\pi/3}\}$, where $\delta\omega\propto
(H_z)_{12}^{2/3}$ is a nonanalytic shift. One can show, however that
these nonanalytic shifts affect the on-shell Boltzmann equation
at the second order in gradients and hence can be neglected \cite{jkpl}.
Finally we note that it is not hard to generalize Eqn.~(\ref{Gra: s2}) and
include the self-energy $\Sigma_R$.

%
%
%
%
\section{On-shell approximation for the Quantum Bolzmann Equation}
\label{sec: On-shell approximation}

The hermitean part of Eqn.~(\ref{KB2b}) defines the quantum Boltzmann 
equation (QBE), and anti-hermitean part a nondynamical constraint equation 
(CE). The QBE is the {\it dynamical master equation\/} whose solutions
specify how the effective phase space of the system is populated, and the
CE is the non-dynamical constraint which singles out the physical solutions. 
In the on-shell limit the QBE (upper signs) and OSE~(\ref{KB2b}) simplify 
to
\begin{eqnarray}
 &- i\Diamond \DBR{G_0^{-1}-\Sigma_R}{G^<}_\mp
  + \DBR{G_0^{-1}-\Sigma_R}{G^<}_\mp &
\nonumber\\
 &=  \frac{1}{2}\NBR{\DBR{\Sigma^>}{G^<}_\pm -  \DBR{\Sigma^<}{G^>}_\pm}
    + \DBR{\Sigma^<}{G_R}_\mp &
\label{QBEm II}
\end{eqnarray}
where $\DBR{a}{b}_\mp=ab\mp ba$. In the following we only consider the QBE.

Eqns. (\ref{QBEm II}) specify the dynamics of the quantum Wigner function 
$G^<$. In the on-shell limit the singular structure of $G^<$ can in general
be factored out as follows
\begin{equation}
iG^<={\cal A}_s n+n^\dagger {\cal A}_s
=\pi{\rm sign}(\omega) \sum \tilde g^i\delta(\omega-\omega_i)
\,,\qquad {\rm for}\quad \Gamma\rightarrow 0\,,
\label{G< decom}
\end{equation}
where $n$ stands for the (regular) matrix of ``occupation numbers.''

Upon integrating the QBE from Eqn.~(\ref{QBEm II}) over positive 
frequencies, we obtain the  following {\em on-shell} quantum Boltzmann 
equation for particles:
\begin{equation}
\partial_t f +
       \frac{1}{2}\{\partial_{\vec k}\,\tilde\omega_d,\partial_{\vec x} f_d\}
       - \frac{1}{2}\{\partial_{\vec x}\,\tilde\omega_d,\partial_{\vec k} f_d\}
       + \frac{i}{\hbar}[\tilde\omega_d,f_d] + Rot\; f_d = Coll[f_d].
\label{f III}
\end{equation}
This is our main result. Analogous equation for antiparticles is obtained 
by integrating over negative frequencies. Here $\tilde\omega_d$ is the 
(diagonal) matrix of the quasiparticle frequencies. The derivative terms, 
including the anticommutators, form the hermitian many-field generalization 
of the usual Boltzmann flow term, and the commutator term is the expected
mixing term familiar from the usual Liouville equation. This term is the source
of the nontrivial quantum coherence effects, and together with some additional
rotational terms included in $Rot \;f$ it leads to potential new sources for
baryogenesis.  The collision term  $Coll[f_d]$ in (\ref{f III}) depends on 
the detailed form of the interactions.  In addition to the usual diagonal 
collision integrals it contains new terms which tend to damp the off-diagonal
elements of $f_d$ to zero. 

Due to the nonlocal character of the SD equation, $f_d$ is related
to $G^{<,>}$ in a nontrivial manner in that there is a derivative correction:
\begin{equation}
f_d = \frac{1}{2\hbar}\left\{\tilde\omega_d,\tilde g_d\right\}
-\frac{i}{4\hbar}
\left[\partial_{\vec k}\,\tilde\omega_d,\partial_{\vec x}\tilde g_d\right]
+\frac{i}{4\hbar}
\left[\partial_{\vec x}\,\tilde\omega_d,\partial_{\vec k}\tilde g_d\right],
\label{f}
\end{equation}
where $\tilde g_d={\cal U}\tilde g{\cal U}^\dagger$ is defined in 
Eqn.~(\ref{G< decom}). This definition is the unique one that renders the QBE
in the simple form of Eqn.~(\ref{f III}).

The basic structure of Eqn.~(\ref{f III}) is basis independent; when rotated
into another basis by an unitary matrix $\cal U$, $f_d$ and $\tilde\omega_d$
become $f={\cal U}^\dagger f_d\,{\cal U}$ and
$\tilde\omega={\cal U}^\dagger\tilde\omega_d\,{\cal U}$, respectively, where
$\tilde\omega$ in general is not diagonal. Further, the detailed form of
${\cal R}ot$ term changes. Physically the ${\cal R}ot$ term arises as a
consequence of the space-time dependent rotation required to bring an 
excitation to the local WKB basis; discussion of its detailed form is beyond
the scope of this talk.

In the adiabatic limit, where the wall width $L$ is much larger than the 
interaction rates, $L\gg 1/\Gamma$, collisions efficiently damp away the 
off-diagonal elements (unless interactions are flavour blind), resulting 
in quantum decoherence. In this case the quantum terms vanish, and one 
obtains the diagonal semiclassical Boltzmann equations for the quasiparticle 
distribution functions, which were studied earlier in refs.\ \cite{jpt2} 
and \cite{cjk1}, with only the classical force $\vec F_{cl} = - 
\partial_{\vec x}\,\omega_k$ and ``spontaneous'' terms to fuel baryogenesis. 
However, we conclude by noting that even when $\Gamma L\gg 1$ is satisfied, 
it may be important to include the non-adiabatic effects, when we are 
interested in phenomena that vanish in the adiabatic limit, as may be the 
case with CP violating effects; the result is then suppressed by 
$(\Gamma L)^{-1}$.
\vskip 0.3cm

%
We thank Dietrich B\"odeker, Felipe Freire, Guy Moore, Emil Mottola, 
Michael Schmidt, and Misha Shaposnikov for many discussions and comments.
%
%
%
\nc{\anap}[3]  {{\it Astron.\ Astrophys.\ }{{\bf #1} {(#2)} {#3}}}
\nc{\ap}[3]    {{\it Ann.\ Phys.\ }{{\bf #1} {(#2)} {#3}}}
\nc{\apj}[3]   {{\it Ap.\ J.\ }{{\bf #1} {(#2)} {#3}}}
\nc{\apjl}[3]  {{\it Ap.\ J.\ Lett.\ }{{\bf #1} {(#2)} {#3}}}
\nc{\app}[3]   {{\it Astropart.\ Phys.\ }{{\bf #1} {(#2)} {#3}}}
\nc{\araa}[3]  {{\it Ann.\ Rev.\ Astron.\ Astrophys.\ }{{\bf #1} {(#2)} {#3}}}
\nc{\arnps}[3] {{\it Ann.\ Rev.\ Nucl.\ and Part.\ Sci.\ }{{\bf #1} {(#2)}
{#3}}}
\nc{\ijmp}[3]  {{\it Int.\ J.\ Mod.\ Phys.\ }{{\bf #1} {(#2)} {#3}}}
\nc{\ijtp}[3]  {{\it Int.\ J.\ Theor.\ Phys.\ }{{\bf #1} {(#2)} {#3}}}
\nc{\jmp}[3]   {{\it J.\ Math.\ Phys.\ }{{\bf #1} {(#2)} {#3}}}
\nc{\mpl}[3]   {{\it Mod.\ Phys.\ Lett.\ }{{\bf #1} {(#2)} {#3}}}
\nc{\nat}[3]   {{\it Nature }{{\bf #1} {(#2)} {#3}}}
\nc{\ncim}[3]  {{\it Nuov.\ Cim.\ }{{\bf #1} {(#2)} {#3}}}
\nc{\np}[3]    {{\it Nucl.\ Phys.\ }{{\bf #1} {(#2)} {#3}}}
\nc{\pr}[3]    {{\it Phys.\ Rev.\ }{{\bf #1} {(#2)} {#3}}}
\nc{\prl}[3]   {{\it Phys.\ Rev.\ Lett.\ }{{\bf #1} {(#2)} {#3}}}
\nc{\pl}[3]    {{\it Phys.\ Lett.\ }{{\bf #1} {(#2)} {#3}}}
\nc{\prep}[3]  {{\it Phys.\ Rep.\ }{{\bf #1} {(#2)} {#3}}}
\nc{\phys}[3]  {{\it Physica\ }{{\bf #1} {(#2)} {#3}}}
\nc{\rmp}[3]   {{\it Rev.\ Mod.\ Phys.\ }{{\bf #1} {(#2)} {#3}}}
\nc{\rpp}[3]   {{\it Rep.\ Prog.\ Phys.\ }{{\bf #1} {(#2)} {#3}}}
\nc{\sjnp}[3]  {{\it Sov.\ J.\ Nucl.\  Phys.\  }{{\bf #1} {(#2)} {#3}}}
\nc{\spjetp}[3]{{\it Sov.\ Phys.\ JETP }{{\bf #1} {(#2)} {#3}}}
\nc{\yf}[3]    {{\it Yad.\ Fiz.\ }{{\bf #1} {(#2)} {#3}}}
\nc{\zetp}[3]  {{\it Zh.\ Eksp.\ Teor.\ Fiz.\ }{{\bf #1} {(#2)} {#3}}}
\nc{\zp}[3]    {{\it Z.\ Phys.\ }{{\bf #1} {(#2)} {#3}}}
\nc{\ibid}[3]  {{\sl ibid.\ }{{\bf #1} {(#2)} {#3}}}
%
%

\end{document}